\documentclass[12pt]{article}
\usepackage{a4wide,epsfig}
\newcommand{\Li}{\mathop{\mathrm{Li}}\nolimits}
\newcommand{\Si}{\mathop{\mathrm{S}}\nolimits}

\catcode`@=11
\newcount\@tempcntc
\def\@citex[#1]#2{\if@filesw\immediate\write\@auxout{\string\citation{#2}}\fi
  \@tempcnta\z@\@tempcntb\m@ne\def\@citea{}\@cite{\@for\@citeb:=#2\do
    {\@ifundefined
       {b@\@citeb}{\@citeo\@tempcntb\m@ne\@citea\def\@citea{,}{\bf ?}\@warning
       {Citation `\@citeb' on page \thepage \space undefined}}%
    {\setbox\z@\hbox{\global\@tempcntc0\csname b@\@citeb\endcsname\relax}%
     \ifnum\@tempcntc=\z@ \@citeo\@tempcntb\m@ne
       \@citea\def\@citea{,}\hbox{\csname b@\@citeb\endcsname}%
     \else
      \advance\@tempcntb\@ne
      \ifnum\@tempcntb=\@tempcntc
      \else\advance\@tempcntb\m@ne\@citeo
      \@tempcnta\@tempcntc\@tempcntb\@tempcntc\fi\fi}}\@citeo}{#1}}
\def\@citeo{\ifnum\@tempcnta>\@tempcntb\else\@citea\def\@citea{,}%
  \ifnum\@tempcnta=\@tempcntb\the\@tempcnta\else
   {\advance\@tempcnta\@ne\ifnum\@tempcnta=\@tempcntb \else \def\@citea{--}\fi
    \advance\@tempcnta\m@ne\the\@tempcnta\@citea\the\@tempcntb}\fi\fi}
\catcode`@=12

\begin{document}

\title{
\vskip-3cm{\baselineskip14pt
\centerline{\normalsize DESY~08--085 \hfill ISSN 0418-9833}
\centerline{\normalsize June 2008\hfill}}
\vskip1.5cm
Orthopositronium lifetime: analytic results in ${\mathcal O}(\alpha)$
and ${\mathcal O}(\alpha^3\ln\alpha)$}

\author{
{\sc Bernd A. Kniehl, Anatoly V. Kotikov\thanks{On leave of absence from
Bogoliubov Laboratory for Theoretical Physics, JINR,
141980 Dubna (Moscow region), Russia.}, Oleg L. Veretin}
\\
{\normalsize II. Institut f\"ur Theoretische Physik, Universit\"at Hamburg,}\\
{\normalsize Luruper Chaussee 149, 22761 Hamburg, Germany}
}

\date{}

\maketitle

\begin{abstract}
We present the ${\mathcal O}(\alpha)$ and ${\mathcal O}(\alpha^3\ln\alpha)$
corrections to the total decay width of orthopositronium in closed analytic
form, in terms of basic transcendental numbers, which can be evaluated
numerically to arbitrary precision.

\medskip

\end{abstract}

\newpage


Quantum electrodynamics (QED), the gauged quantum field theory of the
electromagnetic interaction, has celebrated ground-breaking successes in the
twentieth century.
In fact, its multi-loop predictions for the anomalous magnetic moments of the
electron and the muon were found to agree with highest-precision measurements
within a few parts of $10^{-12}$ and $10^{-10}$, respectively.

Another ultrapure laboratory for high-precision tests of QED is provided by
positronium (Ps), the lightest known atom, being the electromagnetic bound
state of the electron $e^-$ and the positron $e^+$, which was discovered in
the year 1951 \cite{D}.
In fact, thanks to the smallness of the electron mass $m$ relative to
typical hadronic mass scales, its theoretical description is not plagued by
strong-interaction uncertainties and its properties, such as decay widths and
energy levels, can be calculated perturbatively in non-relativistic QED
(NRQED) \cite{Caswell:1985ui}, as expansions in Sommerfeld's fine-structure
constant $\alpha$, with very high precision.

Ps comes in two ground states, $^1S_0$ parapositronium ($p$-Ps) and $^3S_1$
orthopositronium ($o$-Ps), which decay to two and three photons, respectively.
In this Letter, we are concerned with the lifetime of $o$-Ps, which has been
the subject of a vast number of theoretical and experimental investigations.
Its first measurement \cite{Deutsch} was performed later in the year 1951 and
agreed well with its lowest-order (LO) prediction of 1949 \cite{Ore:1949te}.
Its first precision measurement \cite{BH}, of 1968, had to wait nine years to
be compared with the first complete one-loop calculation \cite{Caswell:1976nx},
which came two decades after the analogous calculation for $p$-Ps \cite{HB}
being considerably simpler owing to the two-body final state.
In the year 1987, the Ann Arbor group \cite{michigan1987} published a
measurement that exceeded the best theoretical prediction available then by
more than ten experimental standard deviations.
This so-called $o$-Ps lifetime puzzle triggered an avalanche of both
experimental and theoretical activities, which eventually resulted in what now
appears to be the resolution of this puzzle.
In fact, the 2003 measurements at Ann Arbor \cite{Vallery:2003iz} and Tokyo
\cite{Jinnouchi:2003hr},
\begin{eqnarray}
\Gamma(\mbox{Ann Arbor}) &=&
7.0404(10~\mbox{stat.})(8~\mbox{syst.})~\mu s^{-1},
\nonumber\\
\Gamma(\mbox{Tokyo}) &=&
7.0396(12~\mbox{stat.}) (11~\mbox{syst.})~\mu s^{-1},  
\end{eqnarray}
agree mutually and with the present theoretical prediction,
\begin{equation}
\Gamma(\mbox{theory}) = 7.039979(11)~\mu s^{-1}.
\end{equation}
The latter is evaluated from
\begin{eqnarray}
\Gamma(\mbox{theory})&=&\Gamma_0\left[1 + A \frac{\alpha}{\pi}
+\frac{\alpha^2}{3} \ln\alpha 
+B \left(\frac{\alpha}{\pi}\right)^2\right.
\nonumber\\
&&{}-\left. \frac{3\alpha^3}{2\pi} \ln^2 \alpha 
  + C \frac{\alpha^3}{\pi} \ln \alpha   \right],
\label{Gamma}
\end{eqnarray}
where \cite{Ore:1949te}
\begin{equation}
\Gamma_0 = \frac{2}{9}(\pi^2-9)\frac{m\alpha^6}{\pi}
\end{equation}
is the LO result.
The leading logarithmically enhanced ${\mathcal O}(\alpha^2\ln\alpha)$ and
${\mathcal O}(\alpha^3\ln^2\alpha)$ terms were found in
Refs.~\cite{Caswell:1978vz,Khriplovich:1990eh} and Ref.~\cite{Kar},
respectively.
The coefficients $A=-10.286606(10)$
\cite{Caswell:1976nx,Caswell:1978vz,SH,Adkins:2000fg,Adkins:2005eg},
$B=45.06(26)$ \cite{Adkins:2000fg}, and $C=-5.51702455(23)$
\cite{Kniehl:2000dh} are only available in numerical form so far.
Comprehensive reviews of the present experimental and theoretical status of
$o$-Ps may be found in Ref.~\cite{AFS}.

Given the fundamental importance of Ps for atomic and particle physics, it is
desirable to complete our knowledge of the QED prediction in
Eq.~(\ref{Gamma}).
Since the theoretical uncertainty is presently dominated by the errors in the
numerical evaluations of the coefficients $A$, $B$, and $C$, it is an urgent
task to find them in analytical form, in terms of transcendental numbers,
which can be evaluated with arbitrary precision.
In this Letter, this is achieved for $A$ and $C$.
The case of $B$ is beyond the scope of presently available technology, since
it involves two-loop five-point functions to be integrated over a three-body
phase space. 
The quest for an analytic expression for $A$ is a topic of old vintage:
about 25 years ago, some of the simpler contributions to $A$, due to
self-energy and outer and inner vertex corrections, were obtained analytically
\cite{Stroscio:1982wj}, but further progress then soon came to a grinding halt.
The sustained endeavor of the community to improve the numerical accuracy of
$A$ \cite{Caswell:1976nx,Caswell:1978vz,SH,Adkins:2000fg,Adkins:2005eg} is now
finally brought to a termination.

\begin{figure}[ht]
\begin{center}
\includegraphics[width=0.45\textwidth]{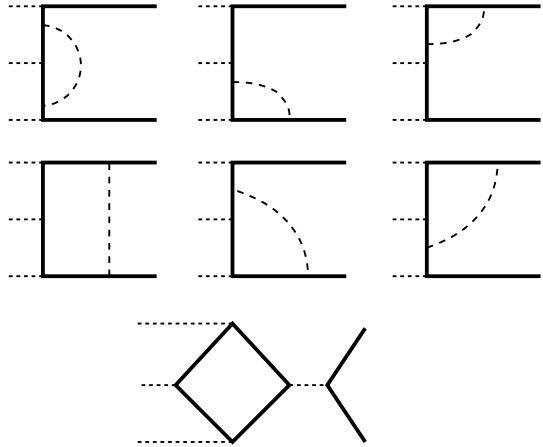}
\end{center}
\caption{\label{fig:dia}%
Feynman diagrams contributing to the total decay width of $o$-Ps at
${\mathcal O}(\alpha)$.
Self-energy diagrams are not shown.
Dashed and solid lines represent photons and electrons, respectively.}
\end{figure}
The ${\mathcal O}(\alpha)$ contribution in Eq.~(\ref{Gamma}),
$\Gamma_1=\Gamma_0A\alpha/\pi$, is due to the Feynman diagrams where a virtual
photon is attached in all possible ways to the tree-level diagrams, with three
real photons linked to an open electron line, and the electron box diagrams
with an $e^+e^-$ annihilation vertex connected to one of the photons being
virtual (see Fig.~\ref{fig:dia}).
Taking the interference with the tree-level diagrams, imposing $e^+e^-$
threshold kinematics, and performing the loop and angular integrations, one
obtains the two-dimensional integral representation \cite{Adkins:2005eg} 
\begin{eqnarray}
\Gamma_1& =& \frac{m\alpha^7}{36\pi^2}
\int\limits^1_0\frac{{\mathrm d}x_1}{x_1}\,\frac{{\mathrm d}x_2}{x_2}\,
\frac{{\mathrm d}x_3}{x_3}\delta(2-x_1-x_2-x_3)
\nonumber\\
&&{}\times[F(x_1,x_3) + {\mathrm{perm.}}],
\label{eq:org}
\end{eqnarray}
where $x_i$, with $0\le x_i\le 1$, is the energy of photon $i$ in the $o$-Ps
rest frame normalized by its maximum value, the delta function ensures energy
conservation, and {\it perm.}\ stands for the other five permutations of
$x_1,x_2,x_3$.
The function $F(x_1,x_3)$ is given by
\begin{equation}
F(x_1,x_3)=g_0(x_1,x_3) + \sum_{i=1}^7 g_i(x_1,x_3) h_i(x_1,x_3),
\end{equation}
where $g_i$ are ratios of polynomials, which are listed in Eqs.~(A5a)--(A5h)
of Ref.~\cite{Adkins:2005eg}, and
\begin{eqnarray}
h_1(x_1) &=& \ln (2x_1),\quad
h_2(x_1)= \sqrt{\frac{x_1}{\overline{x}_1}} \theta_1,
\nonumber\\
h_3(x_1) &=& \frac{1}{2x_1}[\zeta_2-\Li_2(1-2x_1)],
\nonumber\\
h_4(x_1) &=& \frac{1}{4x_1}\left[3\zeta_2-2\theta_1^2\right],\quad
h_5(x_1) = \frac{1}{2\overline{x}_1} \theta_1^2,
\nonumber\\
h_6(x_1,x_3)&=&
\frac{1}{\sqrt{x_1\overline{x}_1x_3\overline{x}_3}}
\left[\Li_2(r^+_A,\overline{\theta}_1)-\Li_2(r^-_A,\overline{\theta}_1)\right],
\nonumber\\
h_7(x_1,x_3)&=&\frac{1}{2\sqrt{x_1\overline{x}_1x_3\overline{x}_3}}
\left[2\Li_2(r^+_B,\theta_1)-2\Li_2(r^-_B,\theta_1)\right.
\nonumber\\
&&{}-\left.\Li_2(r^+_C,0)+\Li_2(r^-_C,0)\right],
\label{1.2}
\end{eqnarray}
with $\overline{x}_i=1-x_i$ and
\begin{eqnarray}
\theta_1 &=& \arctan(\sqrt{\overline{x}_1/x_1}),\quad
\overline{\theta}_1 = \arctan(\sqrt{x_1/\overline{x}_1}),
\nonumber \\
r^{\pm}_A &=& 
\sqrt{\overline{x}_1} \left(1 \pm 
\sqrt{\frac{x_1\overline{x}_3}{\overline{x}_1x_3}} \right),\quad
r^{\pm}_B = \sqrt{x_1} \left(1 \pm 
\sqrt{\frac{\overline{x}_1\overline{x}_3}{x_1x_3}} \right),
\nonumber\\
r^{\pm}_C &=& r^{\pm}_B/\sqrt{x_1}.
\end{eqnarray}
Here, $\zeta_2=\pi^2/6$ and
\begin{equation}
\Li_2(r,\theta) =
-\frac{1}{2}\int\limits^1_0\frac{{\mathrm d}t}{t}
\ln(1-2 r t\cos\theta+r^2 t^2)
 \label{1.4}
\end{equation}
is the real part of the dilogarithm [see line below Eq.~(\ref{eq:s})] of
complex argument $z=r{\mathrm e}^{{\mathrm i}\theta}$ \cite{Lewin}.
Since we are dealing here with a single-scale problem, Eq.~(\ref{eq:org})
yields just one number.

Although Bose symmetry is manifest in Eq.~(\ref{eq:org}), its evaluation is
complicated by the fact that, for a given order of integration, individual
permutations yield divergent integrals, which have to cancel in their
combination.
In order to avoid such a proliferation of terms, we introduce a regularization
parameter, $\delta$, in such a way that the symmetry unter
$x_i\leftrightarrow x_j$ for any pair $i\ne j$ is retained.
In this way, Eq.~(\ref{eq:org}) collapses to
\begin{equation}
\Gamma_1 = \frac{m\alpha^7}{6\pi^2}
\int\limits^{1-\delta}_{2\delta} {\mathrm d}x_{1}
\int\limits^{1-\delta}_{1-x_1+\delta}
\frac{{\mathrm d}x_2}{x_1x_2x_3}F(x_1,x_3),
\label{a.1}
\end{equation}
where $x_3=2-x_1-x_2$.
Note that we may now exploit the freedom to choose any pair of variables $x_i$
and $x_j$ $(i\neq j)$ as the arguments of $F$ and as the integration
variables.

The analytical integration of Eq.~(\ref{a.1}) is rather tedious and requires
a number of tricks to be conceived of.
For lack of space, we can only outline here a few examples.
Specifically, we consider the last two functions of Eq.~(\ref{1.2}), which are
most complicated.
Using Eq.~(\ref{1.4}) and after some manipulations, we obtain the following
integral representation for $h_7(x_1,x_3)$:
\begin{eqnarray}
h_7(x_1,x_3)&=& -\frac{1}{4} \int\limits^1_0 \frac{{\mathrm d}t}{\sqrt{t}
(x_1x_3-\overline{x}_1\overline{x}_3 t)}
\left[\ln\frac{\overline{x}_1x_3}{x_1\overline{x}_3}\right.
\nonumber\\
&&{}+\left.\vphantom{\frac{\overline{x}_1x_3}{x_1\overline{x}_3}}
2 \ln(x_3+\overline{x}_3 t) -\ln t \right].
\end{eqnarray}
Exploiting the $x_1 \leftrightarrow x_3$ symmetry of the coefficient
$g_7(x_1,x_3)$ multiplying $h_7(x_1,x_3)$, this can be simplified as
\begin{eqnarray}
h_7(x_1,x_3)&=& -\frac{1}{4} \int\limits^1_0 \frac{{\mathrm d}t}
{\sqrt{t}(x_1x_3-\overline{x}_1\overline{x}_3 t)}[2 \ln(x_3+\overline{x}_3 t)
\nonumber\\
&&{}-\ln t].
\label{1.6}
\end{eqnarray}
At this point, it is useful to change the order of integrations.
Observing that the logarithmic terms in Eq.~(\ref{1.6}) are $x_1$ independent,
we first integrate over $x_1$ (for a similar approach, see
Ref.~\cite{Kniehl:2005yc}).
In order to avoid the appearance of complicated functions in the intermediate
results, the integration over $t$ in Eq.~(\ref{1.6}) is performed last.

Analogously, $h_6(x_1,x_3)$ can be rewritten as
\begin{eqnarray}
h_6(x_1,x_3)&=& -\frac{1}{2} \int\limits^1_0 \frac{{\mathrm d}t}{\sqrt{t}
(\overline{x}_1x_3-x_1\overline{x}_3 t)}[ \ln x_1 - \ln x_3
\nonumber\\
&&{}+\ln(x_3+\overline{x}_3 t)],
\end{eqnarray}
in which the part proportional to $\ln x_1$ and the complementary part are
first integrated over $x_3$ and $x_1$, respectively.
The $t$ integration is again performed last.

Let us now consider a typical integral that arises upon the first integration:
\begin{eqnarray}
   I = \int\limits_0^1\frac{{\mathrm d}t}{t}\int\limits_0^1
\frac{{\mathrm d}x}{x} \ln[1-4t(1-t)(1-x)]\ln(1-x).
\end{eqnarray}
Direct integration over $t$ or $x$ would lead to rather complicated functions
in the remaining variable.
Instead, we Taylor expand the first logarithm using
$\ln(1-x)=-\sum_{n=1}^\infty x^n/n$ to obtain
\begin{equation}
I= - \sum\limits_{n=1}^\infty \frac{4^n}{n} 
         \int\limits_0^1 \frac{{\mathrm d}t}{t} [t(1-t)]^n 
         \int\limits_0^1 \frac{{\mathrm d}x}{x}(1-x)^n\ln(1-x).
\end{equation}
Now the two integrals are separated and can be solved in terms of Euler's
Gamma function,
$\Gamma(x)=\int_0^\infty{\mathrm d}t\,{\mathrm e}^{-t}t^{x-1}$.
Using
\begin{equation}
   \int\limits_0^1 \frac{{\mathrm d}x}{x}(1-x)^n\ln(1-x)=  - \psi^\prime(n+1),
\end{equation}
where $\psi(x)= {\mathrm d}\ln\Gamma(x)/{\mathrm d}x$ is the digamma function,
we finally have
\begin{equation}
I = \sum\limits_{n=1}^\infty\frac{4^n}{2n}\,\frac{\Gamma^2(n)}{\Gamma(2n)}
\psi^\prime(n+1).
\end{equation}

Another class of typical integrals yields sums involving digamma functions of
half-integer arguments, {\it e.g.}\
\begin{eqnarray}
J&=&\int\limits_0^1 \frac{{\mathrm d}t}{t}\int\limits_0^1{\mathrm d}x
     \frac{\ln[1+4t(1-t)(1-x)]\ln(1-x)}{x-2}
\nonumber\\
&=&\sum\limits_{n=1}^\infty \frac{(-4)^n}{8n}
\frac{\Gamma^2(n)}{\Gamma(2n)}
\left[\psi^\prime\left(\frac{n+2}{2}\right)
-\psi^\prime\left(\frac{n+1}{2}\right) \right].\qquad
\end{eqnarray}
$I$ and $J$ belong to the class of so-called inverse central binomial sums
\cite{Fleischer:1997bw,Kalmykov:2000qe}, and methods for their summation are
elaborated in Ref.~\cite{Kalmykov:2000qe}.
With their help, $I$ and $J$ can be expressed in terms of known irrational
constants, as
\begin{eqnarray}
I&=&-4\zeta_2l_2^2-\frac{l_2^4}{3}-8\Li_4\left(\frac{1}{2}\right)
+\frac{17}{2}\zeta_4,
\nonumber\\
J&=&-\frac{3}{2}\zeta_2l_2^2+\frac{l_2^4}{4}-3\zeta_2l_2l_r+l_2^2l_r^2
+\frac{11}{12}l_2l_r^3+\frac{47}{288}l_r^4
\nonumber\\
&&{}+4l_2l_r\Li_2(r)+\frac{7}{6}l_r^2\Li_2(r)-6l_2\Li_3(-r)
\nonumber\\
&&{}-2l_r\Li_3(-r)+5l_2\Li_3(r)
+\frac{4}{3}l_r\Li_3(r)+6\Li_4\left(\frac{1}{2}\right)
\nonumber\\
&&{}+4\Li_4(-r)-5\Li_4(r)-\frac{13}{3}l_r\Si_{1,2}(r)
+\frac{2}{3}\Si_{1,2}(r^2)
\nonumber\\
&&{}-4\Si_{2,2}(-r)+5\Si_{2,2}(r)
+\zeta_3l_2+\frac{19}{6}\zeta_3l_r,
\end{eqnarray}
where $r =(\sqrt{2}-1)/(\sqrt{2}+1)$, $l_x=\ln x$,
\begin{equation}
\Si_{n,p}(x)=\frac{(-1)^{n+p-1}}{(n-1)!\,p!}
\int_0^1\frac{{\mathrm d}t}{t}\ln^{n-1}t\ln^p(1-tx)
\label{eq:s}
\end{equation}
is the generalized poly-logarithm, $\Li_n(x)=\Si_{n-1,1}(x)$ is the
poly-logarithm of order $n$, and $\zeta_n=\zeta(n)=\Li_n(1)$, with
$\zeta(x)$ being Riemann's zeta function \cite{Lewin,Devoto:1983tc}.

Unfortunately, not all integrals can be computed so straightforwardly.
In more complicated cases, the integrations are not separated after expansion
into infinite series.
We then rely on the PSLQ algorithm \cite{PSLQ}, which allows one to
reconstruct the representation of a numerical result known to very high
precision in terms of a linear combination of a set of irrational constants
with rational coefficients, if that set is known beforehand.
The experience gained with the explicit solution of the simpler integrals
helps us to exhaust the relevant sets.
In order for PSLQ to work in our applications, the numerical values of the
integrals must be known up to typically 150 decimal figures.

After a laborious calculation, we obtain
\begin{eqnarray}
\frac{2}{9}(\pi^2-9) A&=&
\frac{56}{27} 
  - \frac{901}{216}\zeta_2
  - \frac{11303}{192}\zeta_4
  + \frac{19}{6} l_2
  - \frac{2701}{108}\zeta_2 l_2
\nonumber\\
&&{}
  + \frac{253}{24}\zeta_2 l_2^2
  + \frac{251}{144} l_2^4
  + \frac{913}{64} \zeta_2 l_3^2
  + \frac{83}{256} l_3^4
  - \frac{21}{4} \zeta_2 l_2 l_r
\nonumber\\
&&{}
  - \frac{49}{16} \zeta_2 l_r^2
  + \frac{7}{16} l_2 l_r^3
  + \frac{35}{384} l_r^4
  + \frac{581}{16} \zeta_2 \Li_2\left(\frac{1}{3}\right)
\nonumber\\
&&{}
  - \frac{21}{2} l_2 \Li_3(-r)
  - \frac{7}{2} l_r \Li_3(-r)
  + \frac{63}{4} l_2 \Li_3(r)
\nonumber\\
&&{}
  + \frac{63}{8} l_r \Li_3(r)
  - \frac{249}{32} \Li_4\left(-\frac{1}{3}\right)
  + \frac{249}{16} \Li_4\left(\frac{1}{3}\right)
\nonumber\\
&&{}
  + \frac{251}{6} \Li_4\left(\frac{1}{2}\right)
  + 7 \Li_4(-r) 
  - 7 {\rm S}_{2,2}(-r)
\nonumber\\
&&{}
  - \frac{63}{4} \Li_4(r)
  + \frac{63}{4} \Si_{2,2}(r)
  + \frac{11449}{432} \zeta_3
  - \frac{91}{6} \zeta_3 l_2
\nonumber\\
&&{}
  - \frac{35}{8} \zeta_3 l_r
  + \frac{1}{\sqrt2}\left[ 
    \frac{49}{2} \zeta_2 l_r
  - \frac{7}{72} l_r^3
  - \frac{35}{6} l_r \Li_2(r)
\right.
\nonumber\\
&&{}  +\left. \frac{35}{6} \Li_3(r)
  - \frac{175}{3} \Si_{1,2}(r)
  + \frac{14}{3} \Si_{1,2}(r^2)
      + \frac{119}{3} \zeta_3  \right].
\label{Ares}
\end{eqnarray}

The constant $C$ in Eq.~(\ref{Gamma}) is related to $A$ through
\cite{Kniehl:2000dh}
\begin{equation}
C = \frac{A}{3}- \frac{229}{30} + 8 l_2.
\label{Cres}
\end{equation}

From Eqs.~(\ref{Ares}) and (\ref{Cres}), $A$ and $C$ can be numerically
evaluated with arbitrary precision,
\begin{eqnarray}
A &=& -10.28661\,48086\,28262\,24015\,01692\,10991\,\dots\,,
\nonumber\\
C &=& - 5.51702\,74917\,29858\,27137\,88660\,98665\,\dots\,.\qquad
\end{eqnarray}
These numbers agree with the best existing numerical evaluations
\cite{Adkins:2005eg,Adkins:2000fg} within the quoted errors.

In conclusion, we obtained the ${\mathcal O}(\alpha)$ and
${\mathcal O}(\alpha^3\ln\alpha)$ corrections to the total decay width of
$o$-Ps, {\it i.e.} the coefficients $A$ and $C$ in Eq.~(\ref{Gamma}),
respectively, in closed analytic form.
Another important result is the appearance of new irrational constants in
Eq.~(\ref{Ares}).
These constants enlarge the class of the known constants in single-scale
problems.
The constant $B$ in Eq.~(\ref{Gamma}) still remains analytically unknown.

We are grateful to G.S. Adkins for providing us with the computer code
employed for the numerical analysis in Ref.~\cite{Adkins:2005eg}.
This work was supported in part by BMBF Grant No.\ 05~HT6GUA, DFG Grant No.\
SFB~676, and HGF Grant No.\ NG-VH-008.


\begin{thebibliography}{99}

\bibitem{D}
M.~Deutsch,
Phys.\ Rev.\ {\bf 82}, 455 (1951).

\bibitem{Caswell:1985ui}
  W.~E.~Caswell and G.~P.~Lepage,
  Phys.\ Lett.\  B {\bf 167}, 437 (1986).

\bibitem{Deutsch}
M.~Deutsch,
Phys.\ Rev.\ {\bf 83}, 866 (1951).

\bibitem{Ore:1949te}
  A.~Ore and J.~L.~Powell,
  Phys.\ Rev.\  {\bf 75}, 1696 (1949).

\bibitem{BH}
R.~H.~Beers and V.~W.~Hughes,
Bull.\ Am.\ Phys.\ Soc.\ {\bf 13}, 633 (1968).

\bibitem{Caswell:1976nx}
  W.~E.~Caswell, G.~P.~Lepage, and J.~R.~Sapirstein,
  Phys.\ Rev.\ Lett.\ {\bf 38}, 488 (1977).

\bibitem{HB}
I.~Harris and L.~M.~Brown,
Phys.\ Rev.\ {\bf 105}, 1656 (1957).

\bibitem{michigan1987}
C.~I.~Westbrook, D.~W.~Gidley, R.~S.~Conti, and A.~Rich,
Phys. Rev. Lett. {\bf 58}, 1328 (1987);
{\bf 58}, 2153(E) (1987);
Phys.\ Rev.\ A {\bf 40}, 5489 (1989).

\bibitem{Vallery:2003iz}
  R.~S.~Vallery, P.~W.~Zitzewitz, and D.~W.~Gidley,
  Phys.\ Rev.\ Lett.\ {\bf 90}, 203402 (2003).

\bibitem{Jinnouchi:2003hr}
  O.~Jinnouchi, S.~Asai, and T.~Kobayashi,
  Phys.\ Lett.\  B {\bf 572}, 117 (2003)
  [arXiv:hep-ex/0308030].

\bibitem{Caswell:1978vz}
  W.~E.~Caswell and G.~P.~Lepage,
  Phys.\ Rev.\  A {\bf 20}, 36 (1979).

\bibitem{Khriplovich:1990eh}
  I.~B.~Khriplovich and A.~S.~Yelkhovsky,
  Phys.\ Lett.\  B {\bf 246}, 520 (1990).

\bibitem{Kar}
S. G. Karshenboim,
Sov.\ Phys.\ JETP {\bf76}, 541 (1993)
[Zh.\ Eksp.\ Teor.\ Fiz.\ {\bf103}, 1105 (1993)].

\bibitem{SH}
M.~A.~Stroscio and J.~M.~Holt,
Phys.\ Rev.\ A {\bf 10}, 749 (1974);
  M.~A.~Stroscio,
  Phys.\ Rept.\  {\bf 22}, 215 (1975);
  G.~S.~Adkins,
  Ann.\ Phys.\ (N.Y.) {\bf 146}, 78 (1983);
G.~S.~Adkins, A.~A.~Salahuddin, and K.~E.~Schalm,
Phys.\ Rev.\ A {\bf 45}, 7774 (1992);
G.~S.~Adkins,
Phys.\ Rev.\ Lett.\ {\bf 76}, 4903 (1996).

\bibitem{Adkins:2000fg}
  G.~S.~Adkins, R.~N.~Fell, and J.~R.~Sapirstein,
  Phys.\ Rev.\ Lett.\ {\bf 84}, 5086 (2000)
  [arXiv:hep-ph/0003028];
Phys.\ Rev.\ A {\bf 63}, 032511 (2001).

\bibitem{Adkins:2005eg}
  G.~S.~Adkins,
  Phys.\ Rev.\ A {\bf 72}, 032501 (2005).
  [arXiv:hep-ph/0506213].

\bibitem{Kniehl:2000dh}
  B.~A.~Kniehl and A.~A.~Penin,
  Phys.\ Rev.\ Lett.\ {\bf 85}, 1210 (2000);
  {\bf 85}, 3065(E) (2000)
  [arXiv:hep-ph/0004267];
  R.~J.~Hill and G.~P.~Lepage,
  Phys.\ Rev.\  D {\bf 62}, 111301(R) (2000)
  [arXiv:hep-ph/0003277];
  K.~Melnikov and A.~Yelkhovsky,
{\it ibid.}\ {\bf 62}, 116003 (2000)
  [arXiv:hep-ph/0008099].

\bibitem{AFS}
  G.~S.~Adkins, R.~N.~Fell, and J.~R.~Sapirstein,
  Ann.\ Phys.\ (N.Y.) {\bf 295}, 136 (2002);
  D.~Sillou,
  Int.\ J.\ Mod.\ Phys.\  A {\bf 19}, 3919 (2004);
S.~N.~Gninenko, N.~V.~Krasnikov, V.~A.~Matveev, and A.~Rubbia,
Phys.\ Part.\ Nucl.\ {\bf 37}, 321 (2006).

\bibitem{Stroscio:1982wj}
  M.~A.~Stroscio,
  Phys.\ Rev.\ Lett.\  {\bf 48}, 571 (1982);
G.~S.~Adkins,
Phys.\ Rev.\ A {\bf 27}, 530 (1983);
{\bf 31}, 1250 (1985).

\bibitem{Lewin}
L. Lewin,
{\it Polylogarithms and Associated Functions}
(Elsevier, New York, 1981).

\bibitem{Kniehl:2005yc}
  B.~A.~Kniehl and A.~V.~Kotikov,
  Phys.\ Lett.\  B {\bf 638}, 531 (2006)
  [arXiv:hep-ph/0508238].

\bibitem{Fleischer:1997bw}
  J.~Fleischer, A.~V.~Kotikov, and O.~L.~Veretin,
  Phys.\ Lett.\  B {\bf 417}, 163 (1998)
  [arXiv:hep-ph/9707492];
  Nucl.\ Phys.\ {\bf B547}, 343 (1999)
  [arXiv:hep-ph/9808242];
  A.~I.~Davydychev and M.~Yu.~Kalmykov,
{\it ibid.}\ {\bf B699}, 3 (2004)
  [arXiv:hep-th/0303162];
  B.~A.~Kniehl and A.~V.~Kotikov,
  Phys.\ Lett.\  B {\bf 642}, 68 (2006)
  [arXiv:hep-ph/0607201];
  A.~Kotikov, J.~H.~K\"uhn, and O.~Veretin,
  Nucl.\ Phys.\ {\bf B788}, 47 (2008)
  [arXiv:hep-ph/0703013].

\bibitem{Kalmykov:2000qe}
  M.~Yu.~Kalmykov and O.~Veretin,
  Phys.\ Lett.\  B {\bf 483}, 315 (2000)
  [arXiv:hep-th/0004010].

\bibitem{PSLQ}
  H.~R.~P.~Ferguson and D.~H.~Bailey, RNR Technical Report No.\ RNR-91-032;
  H.~R.~P.~Ferguson, D.~H.~Bailey and S.~Arno, NASA Technical Report No.\ 
  NAS-96-005.

\bibitem{Devoto:1983tc}
  A.~Devoto and D.~W.~Duke,
  Riv.\ Nuovo Cim.\  {\bf 7N6}, 1 (1984).

\end{thebibliography}
\end{document}